\documentclass[aps,prl,showpacs,showkeys,amssymb,twocolumn]{revtex4}

\usepackage{bm}
\usepackage{graphicx}
\usepackage{amsfonts}
\usepackage{amsmath}
\usepackage{amssymb}

\begin{document}
	
\title{A model with symmetry-breaking phase transition triggered by a double-well potential}
	
\author{Fabrizio Baroni}
	
\email{f.baroni@ifac.cnr.it, baronifab@libero.it}
\affiliation{CNR - IFAC 'Nello Carrara' Institute of applied physics, Sesto Fiorentino (FI), Italy}
	
\date{\today}
	
\begin{abstract}
	In some recent papers some theorems on sufficiency conditions for the occurrence of a $\mathbb{Z}_2$-symmetry breaking phase transition ($\mathbb{Z}_2$-SBPT) have been showed making use of geometric-topological concepts of potential energy landscapes. In particular, a $\mathbb{Z}_2$-SBPT can be triggered by double-well potentials, or in an equivalent way, by dumbbell-shaped equipotential surfaces. In this paper we introduce a model with a classical $\mathbb{Z}_2$-SBPT which, due to its essential feature, shows in the clearest way the generating-mechanism of a $\mathbb{Z}_2$-SBPT above mentioned. Despite the model is not a physical model, it has all the features of such a model with the same kind of SBPT. At the end of the paper a comparison with the $\phi^4$ model is made. The model may be useful for didactic purposes.
\end{abstract}
	
	
	
\maketitle

\section{Introduction}

Phase transition (PT) are very common in nature. They are sudden changes of the macroscopic behavior of a natural system composed by many interacting parts occurring while an external parameter is smoothly varied. PTs are an example of emergent behavior, i.e., of a collective properties having no direct counterpart in the dynamics or structure of individual atoms \cite{lebowitz}. The successful description of PTs starting from the properties of the interactions between the components of the system is one of the major achievements of equilibrium statistical mechanics.

From a statistical-mechanical viewpoint, in the canonical ensemble, describing a system at constant temperature $T$, a PT occurs at special $T$-values called transition points, where thermodynamic quantities such as pressure, magnetization, or heat capacity, are non-analytic-$T$ functions; these points are the boundaries between different phases of the system. PTs are strictly related to the phenomenon of spontaneous symmetry breaking (SB). For example, in a natural magnet below the Curie temperature the $0(3)$ symmetry is spontaneous broken witnessed by the occurrence of a non-vanishing spontaneous magnetization. In this paper we mostly consider the origin of this aspect, and secondarily the origin of non-analytic points in the thermodynamic functions.

Despite great achievements in our understanding of PTs, yet, the situation is not completely satisfactory. For example, while necessary conditions for the presence of a PT can be found, nothing general is known about sufficient conditions, apart some particular cases \cite{bc}: no general procedure is at hand to tell if a system where a PT is not ruled out from the beginning does have or not such a transition without computing the partition function $Z$. This might indicate that our deep understanding of this phenomenon is still incomplete. 

These considerations motivate a study of PTs based on alternative approaches. One of them is the geometric-topological approach based on the study of the energy potential landscape. In particular, the equipotential surfaces, i.e., the potential level sets ($v$-level sets), gain a great importance inside this approach. This idea has been discussed and tested in many recent papers \cite{acprz,b1,bc,ccp2,ckn,gss,gm,k,rs}.

In particular, in Ref. \cite{b3} it has been stated a links between the occurrence of a $\mathbb{Z}_2$-SBPT and \emph{dumbbell-shaped} $v$-level sets. Intuitively, a $v$-level set is said dumbbell-shaped when it shows two major components connected by a shrink neck. Something like this SBPT generating-mechanism has been put forward also in Refs. \cite{gfp,gfp1}. According to this framework, a spontaneous $\mathbb{Z}_2$-SB is entailed by dumbbell-shaped $v$-level sets, and the thermodynamic critical potential $\langle v\rangle_c$ is in correspondence of a critical $v$-level set in the sense that it is the boundary between the dumbbell $v$-level sets at $v<\langle v\rangle_c$ and the non-dumbbell ones at $v>\langle v\rangle_c$. An advantage with respect to the traditional definition of PTs is that this definition holds for finite $N$ without non-necessarily resorting to the thermodynamic limit. Since in the last decades many examples of transitional phenomena in systems far form the thermodynamic limit have been found (e.g., in nuclei, atomic clusters, biopolymers, superconductivity, superfluidity), a description of PTs valid also for finite systems would be desirable. 

In this paper we will introduce a model showing a $\mathbb{Z}_2$-SBPT which illustrates, in the clearest way, the generating-mechanism based on the concept of dumbbell-shaped $v$-level sets, generated in turn by double-well potentials. The model does not describe any physical system, so that its usefulness is for giving hints about physical models, and for didactic purposes. In Sec. \ref{framework} we will introduce the framework of the geometric-topological approach to SBPTs in the canonical treatment. In Sec. \ref{R4} we will build the new model with the potential landscape characterized by having three stationary points only.

\section{Framework of the geometric-topological approach to SBPTs}
\label{framework}

Hereafter, we will refer to the canonical treatment, although the dumbbell-shaped $v$-level set approach can be extended to the microcanonical one.

Consider an $N$ degrees of freedom system with Hamiltonian given by
\begin{equation}
H(\textbf{p},\textbf{q})=T+V=\sum_{i=1}^N \frac{p_i^2}{2}+V(\textbf{q}).
\end{equation}
Let $M\subseteq\mathbb{R}^N$ be the configuration space. The partition function is by definition
\begin{eqnarray}
Z(\beta,N)=\int_{\mathbb{R}^N\times M} \rm d\mathbf{p}\,\rm d\mathbf{q}\,e^{-\beta H(\mathbf{p},\mathbf{q})}=\nonumber
\\
=\int_{\mathbb{R}^N} \rm d\mathbf{p}\,e^{-\beta\sum_{i=1}^N \frac{p_i^2}{2}}\int_M \rm d\mathbf{q}\,e^{-\beta V(\textbf{q})}=Z_{kin}Z_c,
\end{eqnarray}
where $\beta=1/T$ (in unit $k_B=1$), $Z_{kin}$ is the kinetic part of $Z$, and $Z_c$ is the configurational part. In order to develop what follows, we assume the potential to be lower bounded, thus 
$Z_c$ can be written according to the coarea formula \cite{fr} as follows 
\begin{eqnarray}
Z_c=N\int_{v_{min}}^{+\infty}\rm dv\,e^{-\beta Nv}\int_{\Sigma_{v,N}}\frac{\rm d\Sigma}{\left\|\nabla V\right\|},
\label{zc}
\end{eqnarray}
where $v=V/N$ is the potential density, and the $\Sigma_{v,N}$'s are the $v$-level sets defined as
\begin{equation}
\Sigma_{v,N}=\{\textbf{q}\in M: v(\textbf{q})=v\}.
\label{sigmav}
\end{equation}
The set of the $\Sigma_{v,N}$'s is a foliation of configuration space $M$ while varying $v$ between $v_{min}$ and $+\infty$. The $\Sigma_{v,N}$'s are very important submanifolds of $M$ because as $N\rightarrow\infty$ the canonical statistic measure shrinks around $\Sigma_{\langle v\rangle(T),N}$, where $\langle v\rangle(T)$ is the average potential density. Thus, $\Sigma_{\langle v\rangle(T),N}$ becomes the most probably accessible $v$-level set by the representative point of the system. This fact may have significant consequences on the symmetries of the system because of the fact that the ergodicity may be broken by the mechanisms pointed out in Refs. \cite{bc,b3}.

We can make the same considerations for $Z_{kin}$, but the related submanifolds $\Sigma_{e,N}$, where $e=E/N$ is the kinetic energy density, are all trivially $N$-spheres, thus they cannot affect the symmetry properties of the system. Furthermore, $Z_{kin}$ is analytic at any $T$ in the thermodynamic limit, so that it cannot entail any loss of analyticity in $Z$. For the considerations above, hereafter we will consider $Z_c$ alone, so as for the thermodynamic functions.

\section{The new model in mean-field version}
\label{R4}

In this section we will introduce a new model that we will call $R^4$ \emph{model}. It shows in the most evident way the generating-mechanism of $\mathbb{Z}$-SBPTs based on dumbbell-shaped $\Sigma_v$'s. 

Consider a central potential with an $O(N)$ symmetry given by the term $|\mathbf{q}|^4=R^4$ multiplied by a suitable constant, where $R^2$ is the square radius $R^2=\sum_{i=1}^N q_i^2$, to which we add the mean-field-Ising-like interacting term 
\begin{equation}
V=\frac{1}{4N}\left(\sum_{i=1}^N q_i^2\right)^2-\frac{J}{2N}\left(\sum_{i=1}^N q_i\right)^2=\frac{1}{4N}R^4-\frac{NJ}{2}m^2.
\label{VR4}
\end{equation}
The factor $1/N$ has been inserted to guarantee $v=V/N$ to be intensive.
We will show that this model undergoes a classical $\mathbb{Z}_2$-SBPT. It belongs to the class of the revolution models defined in Ref. \cite{b4} because the subsets at constant magnetization of the $\Sigma_{v,N}$'s are $(N-1)$-spheres, but with the advantage that the potential can be written in the standard coordinate system. 

Beside the mean-field version, we can consider also the short-range versions of the $R^4$ model. We conjecture that the $\mathbb{Z}_2$-SBPT occurs for any dimension $d$, including the special case $d=1$ with vanishing critical temperature.  

Let us consider the geometric-topological analysis of the potential. $\nabla V=0$ is equivalent to the following system
\begin{equation}
q_i\sum_{j=1}^N q_j^2-J\sum_{j=1}^N q_j=0, \quad i=1,\cdots,N.
\end{equation}
It is easy to show that the solutions $q^s_i$'s for $i=1,\cdots,N$ are all equal and satisfy the following equation
\begin{equation}
{q_i^s}^3-J q_i^s=0,\quad i=1,\cdots,N.,
\end{equation}
whence $q^s_i=\sqrt{J}$, $q^s_i=-\sqrt{J}$  for $i=1,\cdots,N$. Summarizing, there are two global minima and a saddle. The global minimum value of $V$ is 
\begin{equation}
V_{min}=-\frac{1}{4}NJ^2.
\end{equation}
	\begin{figure}
		\begin{center}
		\includegraphics[width=0.235\textwidth]{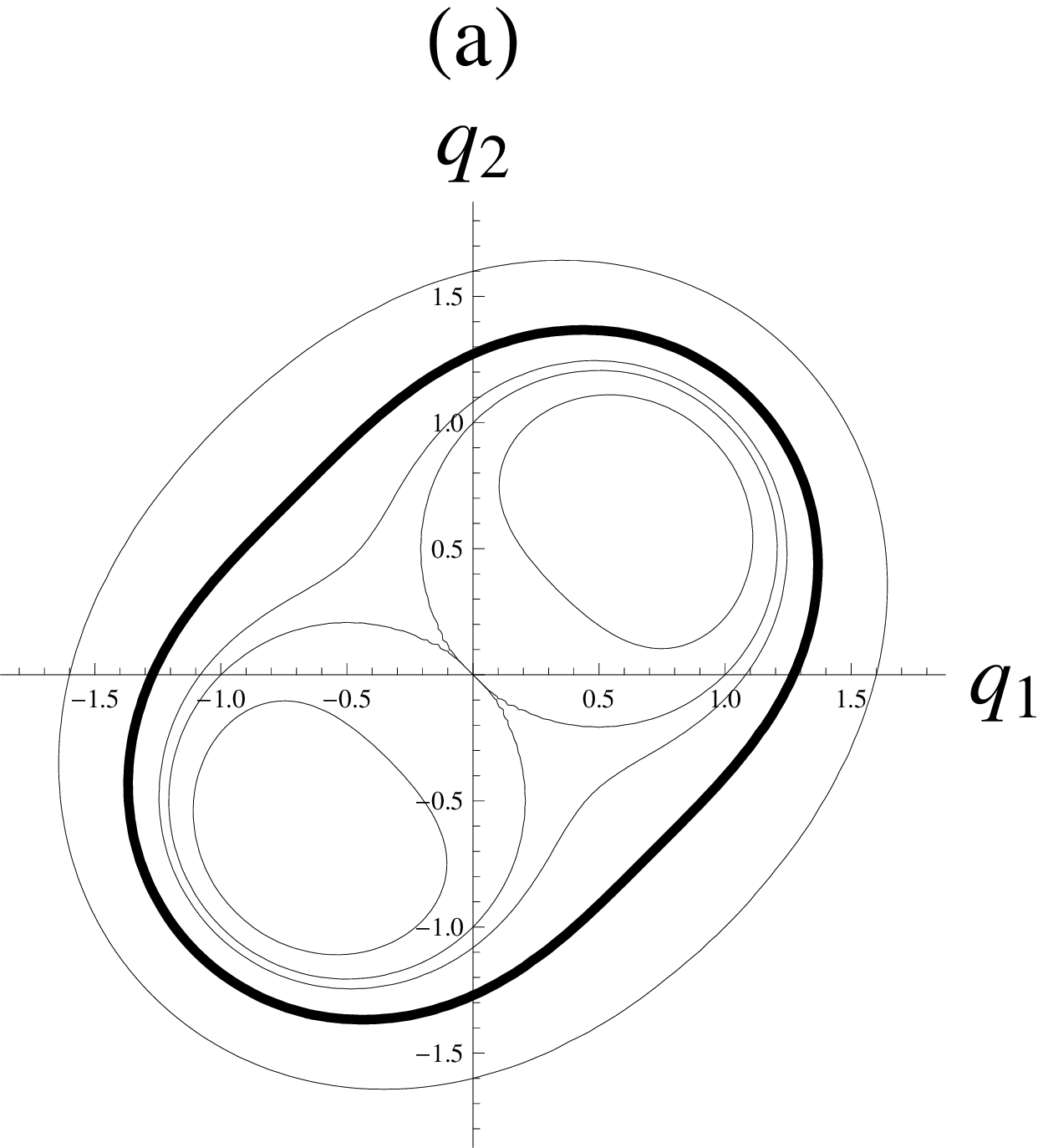}
		\includegraphics[width=0.235\textwidth]{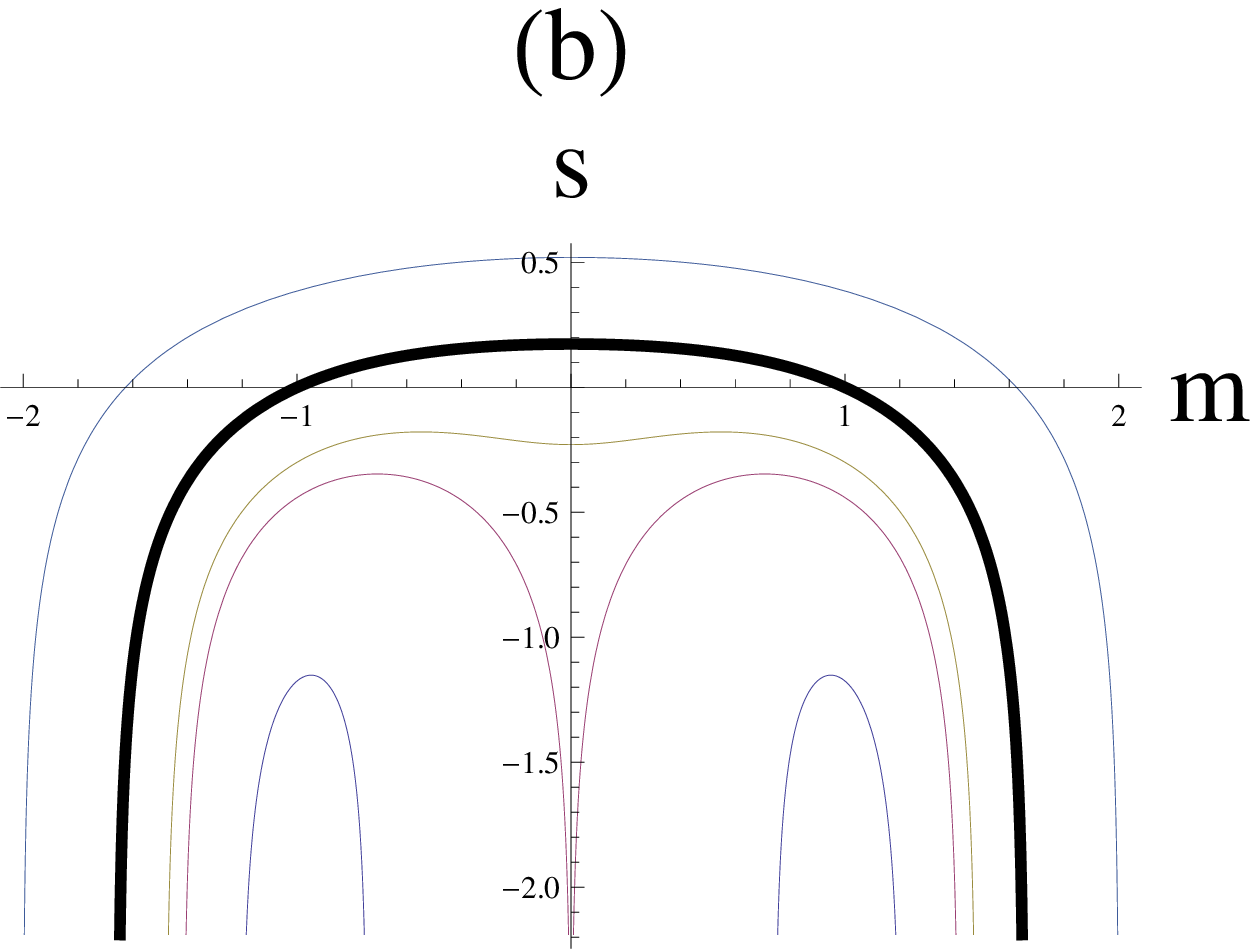}
		\caption{(a) Some $\Sigma_{v,N}$'s for $N=2$ of the model (\ref{VR4}) for $v=-0.4, 0, 0.1 ,0.5, 1$, respectively from the innermost to the outermost. $\Sigma_{0.5,2}$ is marked. In $N$ dimensions it is the boundary between the dumbbell-shaped $\Sigma_{v,N}$'s and the those witch are not. (b) Microcanonical entropy for the same $v$-values, respectively form the lowest graph to the highest one.}
		\label{fig_R4_sigmav_sm}
	\end{center}
\end{figure}

In panel (a) of Fig. \ref{fig_R4_sigmav_sm} some $\Sigma_{v,N}$'s for $N=2$ are represented. Note that $\Sigma_{0,2}$ is made up by the union of two circles ($N$-spheres in $N$ dimensions) which touch each other at the critical point $(0,0)$. It has no particular significance. 

Before entering the analysis of configuration space from the perspective of the dumbbell-shaped $\Sigma_{v,N}$'s, we observe that the potential (\ref{VR4}) satisfies two general sufficiency conditions for $\mathbb{Z}_2$-SBPT to occur: Theorem $1$ in Ref. \cite{bc} and the theorem in Sec. 2.2. in Ref. \cite{b3}. This is essentially due to the fact that the two global minima of the potential are separated by a gap proportional to $N$.

\subsection{Dumbbell-shaped $\Sigma_{v,N}$'s}

\begin{figure}
	\begin{center}
		\includegraphics[width=0.49\textwidth]{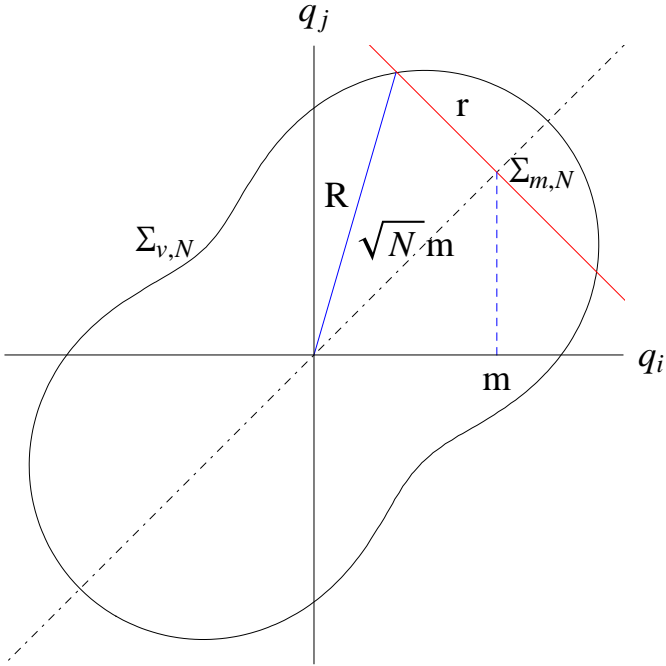}
		\caption{Schematic representation in two dimensions of $\Sigma_{v,N}\cap\Sigma_{m,N}$ made by the intersection between the line  orthogonal to the axis of symmetry of the closed line ($\Sigma_{m,N}$) and the closed line ($\Sigma_{v,N}$). It is an $(N-1)$-sphere whose radius is $r$.}
		\label{R4_sketch}
	\end{center}
\end{figure}

$\Sigma_{v,N}\cap\Sigma_{m,N}$ is an $(N-1)$-sphere whose radius $r$ is linked to $R$ and $m$ via the Phytagorian theorem and the definition (\ref{VR4}) (Fig. \ref{R4_sketch} can help the reader)
\begin{equation}
r(v,m)=\sqrt{N}\left(2\left(v+\frac{J}{2}m^2\right)^{\frac{1}{2}}-m^2\right)^{\frac{1}{2}}.
\end{equation}
The microcanonical volume is
\begin{equation}
\omega_N(v,m)=\int_{\Sigma_{v,N}\cap\Sigma_{m,N}}\frac{d\Sigma}{\|\nabla V\wedge \nabla M\|},
\end{equation}
where
\begin{equation}
\Vert\nabla V\wedge\nabla M\Vert^2=\det\left(
\begin{matrix}
\nabla V\cdot\nabla V & \nabla M\cdot\nabla V
\\
\nabla V\cdot\nabla M & \nabla M\cdot\nabla M
\end{matrix}
\right)
\end{equation}
is the determinant of the Gramian of $V$ and $M=Nm$. By some trivial algebraic manipulation we can show that it is a polynomial in $N$, $v$ and $m$ such that $\Vert\nabla V\wedge\nabla M\Vert^{1/N}\to 1$ for $N\to\infty$. Since the Gramian was able to pass out of the integral sign because of the $O(N-1)$ symmetry of the model, the microcanonical entropy becomes
\begin{equation}
\omega_N(v,m)= \frac{C(N-1)}{\Vert\nabla V\wedge\nabla M\Vert}r^{N-2},
\end{equation}
where $C(N-1)=2\pi^{(N-1)/2}/\Gamma\left((N-1)/2\right)$ is the volume of the unitary  $(N-1)$-sphere. Finally, in the thermodynamic limit $N\rightarrow\infty$, the microcanonical entropy $s_N(v,m)=\ln \omega_N(v,m)^{1/N}$ becomes
\begin{equation}
s(v,m)=\frac{1}{2}\ln\left(2\left(v+\frac{J}{2}m^2\right)^{\frac{1}{2}}-m^2\right).
\end{equation}

Now, our purpose is to find out the analytic relation between the spontaneous magnetization and the average specific potential, and in particular the critical average potential $\langle v\rangle_c$, by studying $s(v,m)$ as a function of $m$. Indeed, according to the theorem in Sec. 2.2 in Ref. \cite{b3}, the $\Sigma_{\langle v\rangle_c,N}$ is the boundary between the dumbbell-shaped $\Sigma_{v,N}$'s and those which are not. At fixed $v$, $\langle m\rangle$ is related to the $\Sigma_{v,N}\cap\Sigma_{m,N}$ of maximum volume, i.e., when $\partial s/\partial m=0$. After some trivial algebraic manipulation, we get

 \begin{equation}
\frac{\partial s}{\partial m}=\frac{1}{2}e^{-2s(v,m)}\left(J\left(v+\frac{J}{2}m^2\right)^{-\frac{1}{2}}-2\right)m=0,
\end{equation}
hence the solution

\begin{equation}
\langle m\rangle(v)=\begin{cases}
 \pm \left(\frac{J}{2}-\frac{2 v}{J}\right)^{\frac{1}{2}}& \text{ if }\,\,\, v_{min}<v\leq \langle v\rangle_c
\\ 
 0& \text{ if } \,\,\,v\geq \langle v\rangle_c
\end{cases},
\label{R4_mv}
\end{equation}
where $\langle v\rangle_c=J^2/4$ (see Fig. \ref{R4_mvT}). Another way for finding out $\langle v\rangle_c$ is setting to zero $\partial^2 s/\partial m^2$ at $m=0$, which after some trivial algebraic manipulation writes as
\begin{equation}
\frac{\partial^2 s}{\partial m^2}\bigg|_{m=0}=\frac{1}{2}v^{-\frac{1}{4}}\left(\frac{J}{\sqrt{v}}-2\right)=0,
\end{equation}
whence the solution.

\subsection{Canonical thermodynamic}

We cannot provide any analytical solution for this model. The presence of the $\mathbb{Z}_2$-SB is guaranteed by Theorem $1$ in Ref. \cite{bc} and the complete $\mathbb{Z}_2$-SBPT by the theorem in Sec. 2.2 in Ref. \cite{b3}. If we consider also the short-range versions of the model, the dimension $d$ of the lattice enters the game. We conjecture that the model undergoes the $\mathbb{Z}_2$-SBPT for any $d>1$. $d=1$ is excluded because the gap of the potential density $v$ between the two wells tends to zero as $N\rightarrow\infty$. For more precision, the $\mathbb{Z}_2$-SBPT occurs the same, but the critical temperature is vanishing.

\begin{figure}
	\begin{center}
		\includegraphics[width=0.235\textwidth]{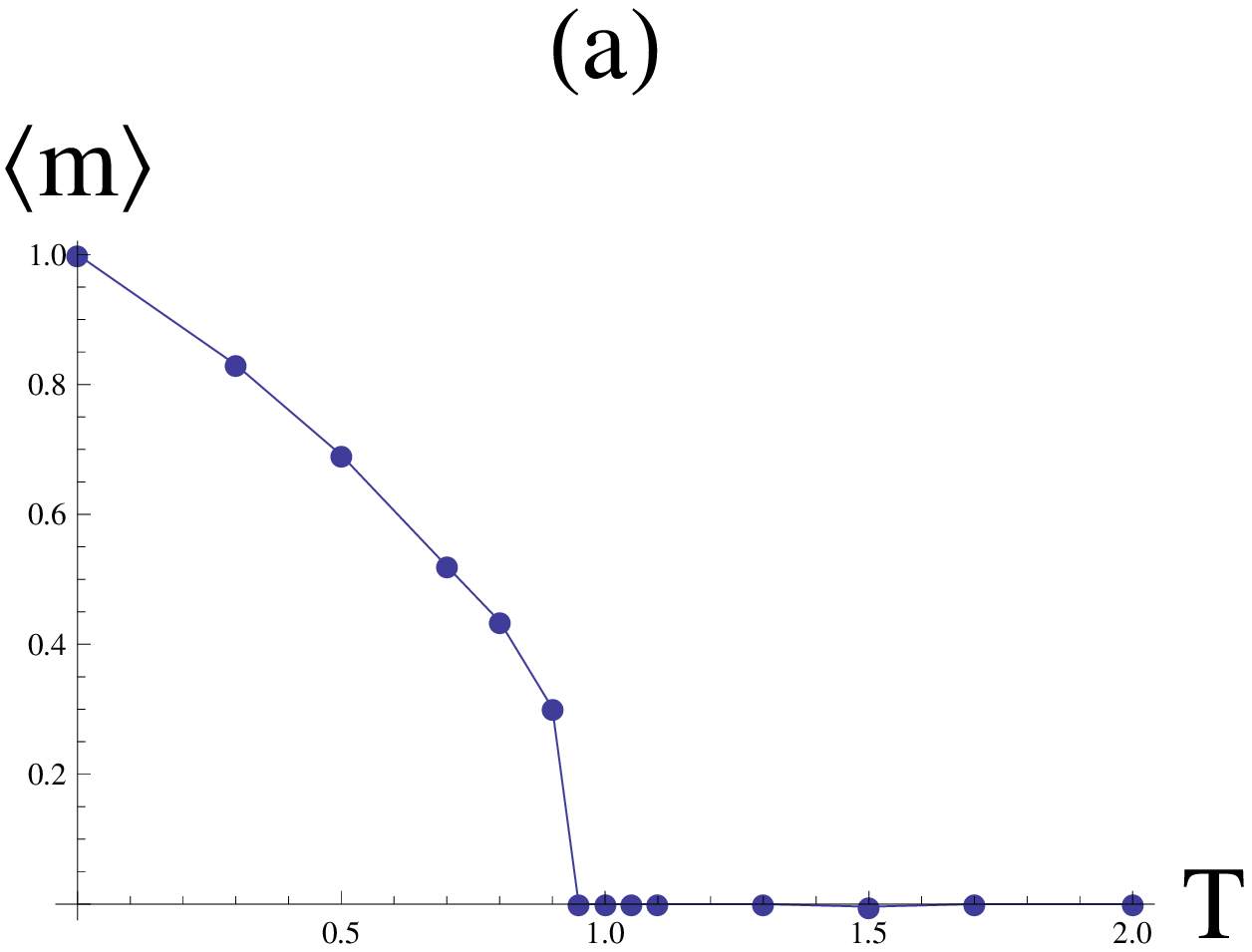}
		\includegraphics[width=0.235\textwidth]{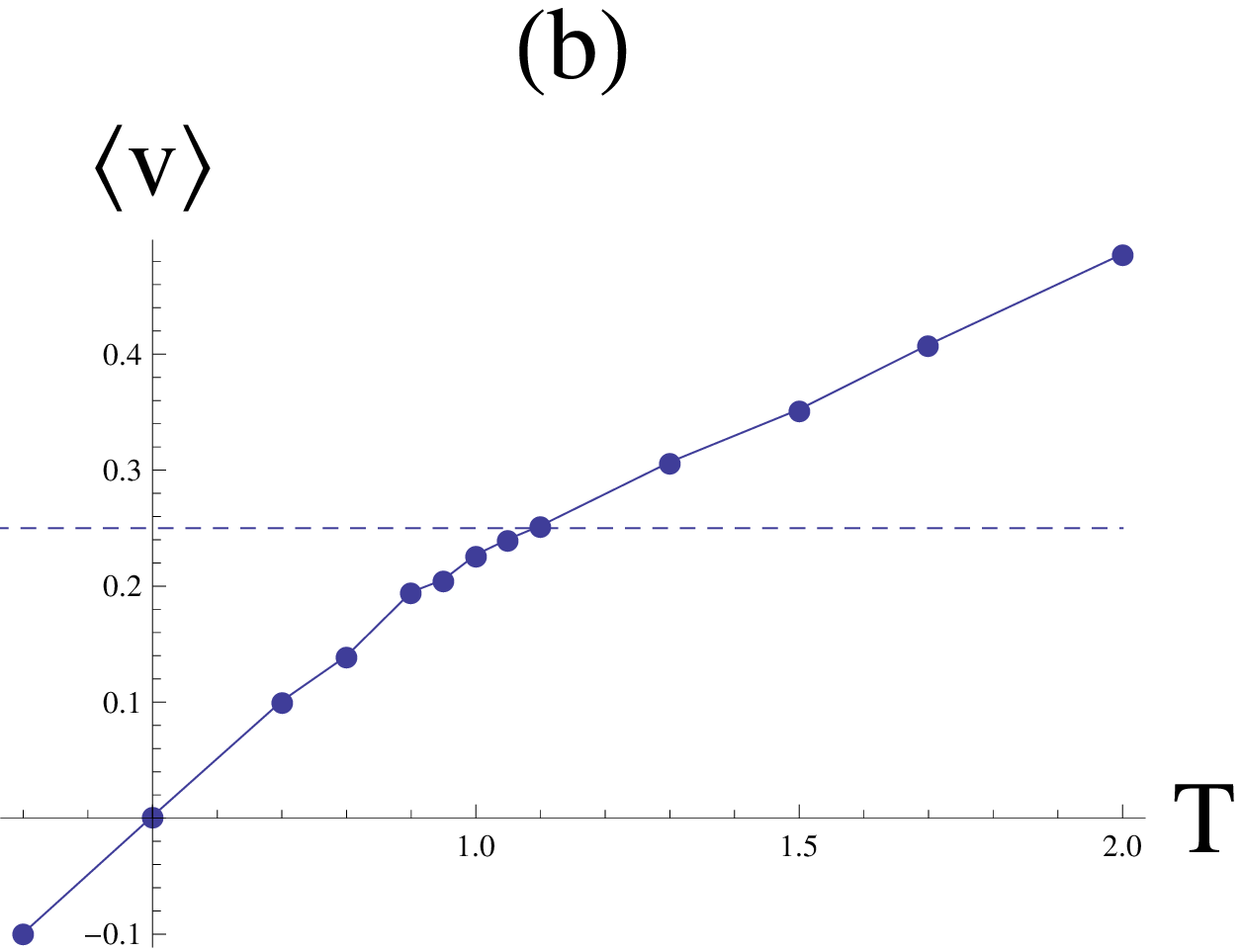}
		\includegraphics[width=0.235\textwidth]{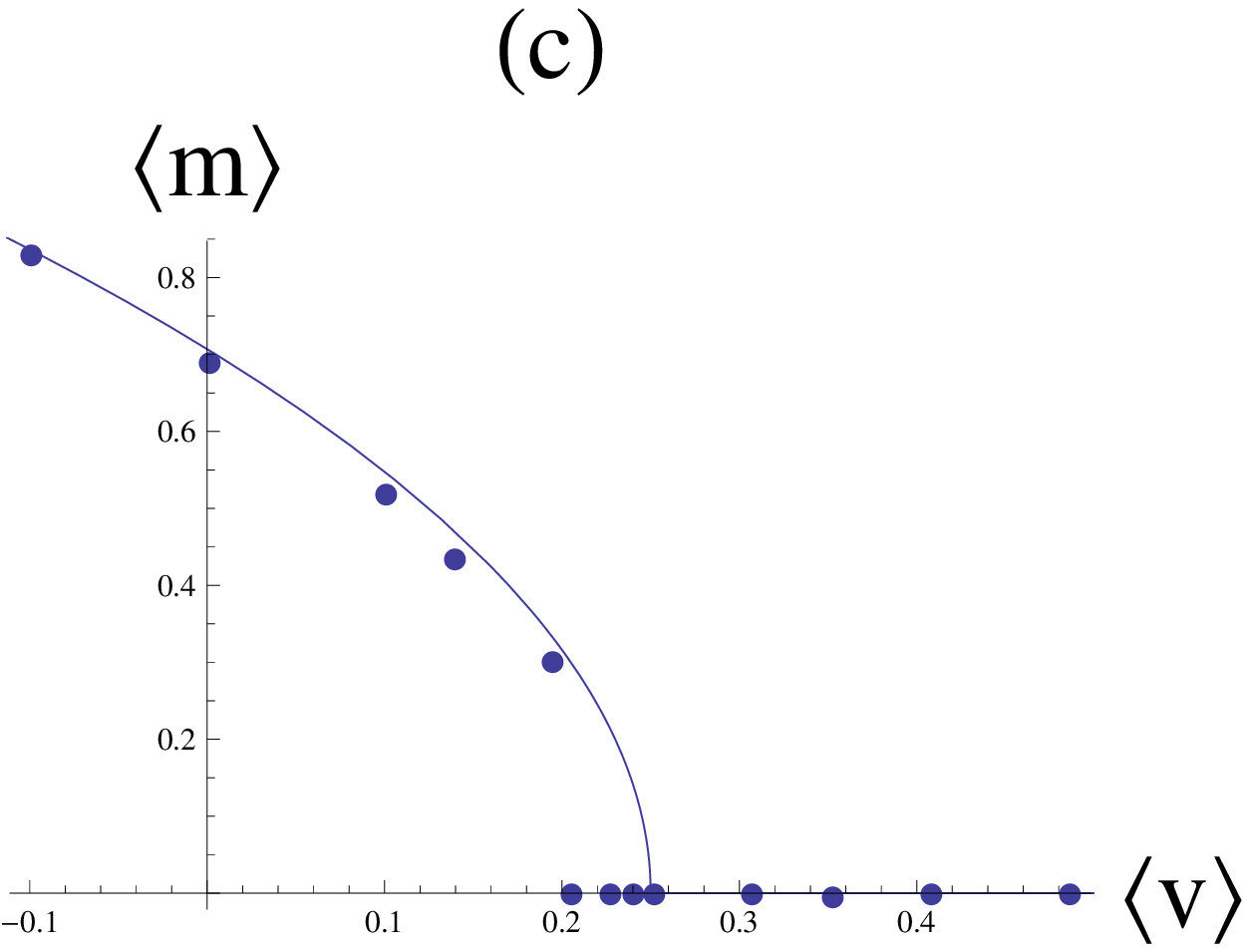}
		\caption{(a) Spontaneous magnetization as a function of the temperature of the model (\ref{VR4}) by a Monte Carlo simulation for $N=100$ and $J=1$. (b) As panel (a) for the specific potential. (c) The same result in the (v,m) graph. In panel (a) and (b) the line is a guide for the eye, while in panel (c) it is eq. (\ref{R4_mv}).}
		\label{R4_mvT}
	\end{center}
\end{figure}

The presence of the $\mathbb{Z}_2$-SBPT has been showed by a Monte Carlo simulation. The results are represented in Fig. \ref{R4_mvT}. We note that the critical potential coincides with that calculated geometrically $v_c=1/2$ for $J=1$ with good approximation. This is a confirmation of the validity of the geometric approach of the $\mathbb{Z}$-SBPT based on dumbbell-shaped $\Sigma_v$'s.

\subsection{$R^4$ model versus $\phi^4$ model}

The potential of the on-lattice $\phi^4$ model potential is given by
\begin{equation}
V=V_{conf}-J\sum_{\langle i,j\rangle}q_i q_j,
\label{phi4pot}
\end{equation}
where $J>0$ is the coupling constant and $V_{conf}$ is the confining part of the potential given in turn by
\begin{equation}
V_{conf}=\sum_{i=1}^N \left(\alpha q_i^4-\beta q_i^2\right),
\label{phi4local}
\end{equation}
where $\alpha$ and $\beta$ are positive real constants. The interacting term is the same as that of the $R^4$ model in the mean-field case. Sometimes $V_{comf}$ is called local potential because it is expressed by a summation of terms concerning each one a single degree of freedom. This does not hold for the $R^4$ model. This is a complication that is paid to simplify the shape of the $\Sigma_v$'s from a geometric point of view. Indeed, in the $R^4$ model without the interacting term they are simply $(N-1)$-spheres. In our opinion, there are no other differences between the two models. In particular, in Ref. \cite{b5} the $\phi^4$ model without quadratic term in the local potential has been studied showing a scenario of the $\Sigma_v$'s equivalent at all form the geometric-topological viewpoint to that of the $R^4$ model.

\section{Conclusions}

In this paper we have introduced an Hamiltonian models with continuous $\mathbb{Z}_2$-SBPT entailed by some sufficiency conditions given on the potential energy landscape. These conditions are specified in Ref. \cite{b2}. The substance feature is a double-well potential with gap proportional to $N$, alongside other minor details. A similar but weaker sufficiency condition had already been given in Ref. \cite{bc}

In Ref. \cite{b3} it has been given a straightforward theorem (Theorem 1 in the paper) according to which dumbbell-shaped $\Sigma_{v,N}$'s are necessary and sufficiency condition to entail a $\mathbb{Z}_2$-SBPT. Roughly speaking, a $\Sigma_{v,N}$ is dumbbell-shaped if it is made up by too major components connected by a shrink neck. Generally, this kind of $\Sigma_{v,N}$ stems from a double-well potential. In this framework the thermodynamic critical potential $\langle v\rangle_c$ results to be exactly in correspondence of a critical $\Sigma_{v_c,N}$. The last is the boundary between the dumbbell-shaped $\Sigma_{v_c,N}$'s for $v<v_c$ of the broken phase and those which are not for $v>v_c$ of the unbroken phase. 

Despite the model here introduced is not a physical model, at least as far as we can see
now, it has all the features of a such a model undergoing a continuous $\mathbb{Z}_2$-SBPT. The model shows in the clearest way the generating-mechanism of the $\mathbb{Z}_2$-SBPT based on the dumbbell-shaped $\Sigma_{v,N}$'s framework. The model may be suitable for didactic purposes and for numerical and analytical studies. In particular, it is possible the complete analytical study of the curvature properties of the $\Sigma_{v,N}$'s, e.g., by the Gaussian curvature, to be related to the $\mathbb{Z}_2$-SBPT.

\begin{acknowledgments}
	I would like to thank ................
\end{acknowledgments}

\end{document}